\documentclass{article}

\usepackage{PRIMEarxiv}

\usepackage[utf8]{inputenc} 
\usepackage[T1]{fontenc}    
\usepackage{hyperref}       
\usepackage{booktabs}       
\usepackage{amsfonts}       
\usepackage{nicefrac}       
\usepackage{microtype}      
\usepackage{lipsum}
\usepackage{fancyhdr}       
\usepackage{graphicx}       
\graphicspath{{media/}}     
\usepackage{multirow}
\usepackage{graphicx}
\usepackage{color}
\usepackage{xcolor}
\usepackage{caption}
\usepackage{subcaption}
\usepackage{xcolor,graphicx,soul}
\usepackage{amsmath}
\usepackage{amssymb}
\usepackage{booktabs}
\usepackage{multirow}
\usepackage{algorithm}
\usepackage{algpseudocode}
\usepackage{url}
\title{Are Vision xLSTM Embedded $U$-Nets More Reliable in Medical 3D Image Segmentation?}

\author{
  Pallabi Dutta \\
  Machine Intelligence Unit \\
  Indian Statistical Institute, Kolkata 700108, India\\
  \texttt{duttapallabi\_r@isical.ac.in} \\
  \And
  Soham Bose \\
  Department of Computer Science and Engineering \\
  Jadavpur University,
  Kolkata 700032, India\\
  \texttt{sohamb.cse.ug@jadavpuruniversity.in} \\
  \AND
  Swalpa Kumar Roy \\
  Department of Computer Science and Engineering \\
  Alipurduar Government Engineering \\
  and Management College, West Bengal 736206, India\\
  \texttt{swalpa@agemc.ac.in} \\
  \And
  Sushmita Mitra \\
  Machine Intelligence Unit \\
  Indian Statistical Institute, Kolkata 700108, India\\
  \texttt{sushmita@isical.ac.in} 
}

\begin{document}
\maketitle

\begin{abstract}
The development of efficient segmentation strategies for medical images has evolved from its initial dependence on Convolutional Neural Networks (CNNs) to the current investigation of hybrid models that combine CNNs with Vision Transformers (ViTs). There is an increasing focus on creating architectures that are both high-performing and computationally efficient, capable of being deployed on remote systems with limited resources. Although transformers can capture global dependencies in the input space, they face challenges from the corresponding high computational and storage expenses involved. This research investigates the integration of CNNs with Vision Extended Long Short-Term Memory ({\it Vision-xLSTM})s  by introducing the novel {\it U-VixLSTM}. 

The {\it Vision-xLSTM} blocks capture the temporal and global relationships within the patches extracted from the CNN feature maps. The convolutional feature reconstruction path upsamples the output volume from the {\it Vision-xLSTM} blocks to produce the segmentation output. Our primary objective is to propose that {\it Vision-xLSTM} forms an appropriate backbone for medical image segmentation, offering excellent performance with reduced computational costs. The {\it U-VixLSTM} exhibits superior performance compared to the state-of-the-art networks in the publicly available {\it Synapse, ISIC} and {\it ACDC} datasets. Code provided: \url{https://github.com/duttapallabi2907/U-VixLSTM}
\end{abstract}

\section{Introduction}

Artificial intelligence-driven segmentation of medical images helps in the diagnosis, treatment planning, and monitoring of patients in different imaging modalities. There are several challenges in accurately defining target regions from medical image data due to their complex characteristics and variations in anatomical structures. Some common issues include ambiguous boundaries of target organs and pathologies, low contrast, overlapping structures, and inter- and intra-patient variability. Analyzing volumetric medical image data requires a substantial amount of computational resources \cite{choi2020development}. These factors lead to the need to develop segmentation algorithms with a high degree of precision and accuracy in the output. Deep learning is widely used in automated medical image segmentation, mainly due to its ability to extract underlying characteristics from input images with minimal human involvement and generalization to unseen samples \cite{chen2022recent}. CNNs and ViTs are two significant deep learning frameworks used in the design of algorithms for medical image segmentation. 

The $U-$Net architecture \cite{ronneberger2015u}, consisting of symmetric encoder-decoder structures with CNNs, revolutionized the extensive use of CNNs in medical image segmentation. Hierarchical modeling of complex high-level patterns from low-level features and the integration of fine-grained spatial information with coarser abstract features are the factors that are attributed to the success of the $U-$Net. Various well-known architectures have been developed to expand on their accomplishments. These include $U-$Net++ \cite{zhou2018unet++}, $V-$Net \cite{milletari2016v}, Attention $U-$Net \cite{oktay2018attention}, $U-$Net 3+ \cite{huang2020unet} and LB-UNet \cite{xu2024lbunet}. Each introduces distinct modifications to improve performance and address specific difficulties inherent in medical image segmentation.

However, despite the numerous advantages of CNNs, they have inherent limitations. CNNs are incapable of capturing the global context of target anatomical structures with varying sizes due to their limited receptive field. In this context, ViTs became increasingly popular for their ability to capture long-range dependencies between different parts of the input image. This led to exploring combinations of CNNs and Transformers to utilize their generic benefits in creating a robust representation of target structures.  ViTs can efficiently learn the overall structural details. In contrast, CNNs focus mainly on local patterns because of their limited field of view \cite{park2021vision}. 

TransUNet \cite{chen2021transunet} pioneered the study of the impact of such a hybrid structure on medical image segmentation. It modeled the global structural dependencies in the intermediate feature volumes, produced from CNN layers, using the self-attention mechanism \cite{vaswani2017attention} of ViT blocks. This framework was also adopted into multiple segmentation tasks in the medical domain with the necessary modifications \cite{cheng2022fully}, \cite{pemmaraju2023cascaded}, \cite{wang2022multiscale} and \cite{zhou2025bsa}.  UNETR \cite{hatamizadeh2022unetr}, TransAttUNet \cite{chen2023transattunet} and UCTransNet \cite{wang2022uctransnet} are some of the other well-known hybrid CNN-Transformer architectures that integrate CNNs with standard ViT blocks.

The self-attention mechanism, the driving force behind Transformers, suffers from the curse of quadratic complexity. This leads to a huge computational burden and memory requirement. Additionally, the fixed-scale patch processing by ViT blocks across multiple levels lacked multi-scale feature learning capability. This is crucial for efficiently segmenting anatomical structures exhibiting variations in shape and size. Swin Transformers \cite{liu2021swin} addresses these drawbacks by initially computing self-attention among the patches within non-overlapping windows rather than the entire input. Patch merging layers at every level led to the hierarchical processing of feature maps.

Swin Transformer blocks were leveraged in multiple medical image segmentation models, {\it viz.} Swin UNETR \cite{hatamizadeh2021swin} and DS-TransUNet \cite{lin2022ds}. The idea of computing self-attention among patches within a local window was also adopted in \cite{wang2022mixed}. A global token representative of each window was derived and fed to a Gaussian-weighted axial attention module to compute cross-dependencies between these representative tokens. The overall computational complexity got reduced to $\mathcal{O}(n\sqrt{n})$, where $n$ is the total patches. Several other models were developed to reduce the quadratic computational complexity of the self-attention mechanism. The efficient attention mechanism \cite{shen2021efficient} was adopted in \cite{azad2023laplacian} to reduce the computational complexity of the self-attention mechanism to $\mathcal{O}(d^2n)$, where $d$ is the embedding dimension. 

 The Transformers still need huge memory requirements despite these advancements in designing the self-attention mechanism while computing the long-range dependencies. This restricts their deployment in resource-constrained environments. Recently, state-space models like Mamba \cite{gu2024mamba} have been leveraged in developing models for medical image segmentation to address these shortcomings, {\it viz.} {\it U-}Mamba \cite{ma2024umamba}, Swin-UMamba \cite{liu2024SwinUMamba} and Swin-UMamba$\dagger$ \cite{liu2024swin+}. Mamba uses state space models to filter relevant information and organize it in a structured memory instead of tracking all tokens, as seen in Transformers. This leads to a linear computational and memory complexity, making it faster than Transformers.

Mamba-based models lack retaining fine-grained details despite efficient computational and memory complexities while updating the structured memory \cite{merill2024the}.  This results from the selective structured memory update mechanism that might lead to information loss. Consequently, it leads to inefficient segmentation performance of small and localized anatomical structures. Additionally, Mamba faces difficulties in refining distant past information, which leads to inefficient long-range context correction.

Extended Long-Short Term Memory (xLSTM) \cite{beck2024xlstm} has recently emerged as a strong contender to Transformers and Mamba in sequence modeling. It addresses the key limitations of Transformers with its inherent linear computational complexity [$\mathcal{O}(n)$] and constant memory complexity [$\mathcal{O}(1)$] with respect to the length of input sequences.  xLSTM extends standard LSTM by incorporating memory structures with higher capacity, in contrast to the scalar memory state of LSTM and compressed state in Mamba. The exponential gating mechanism enhances the ability of the model to revise past information. Vision-xLSTM (ViL) \cite{alkin2024vision} pioneers the application of xLSTMs to computer vision tasks. 

Clinical scenarios with limited computing resources require models that can accurately delineate target structures from medical data while constraining computational power and memory requirements. This is essential to guarantee sustainable and cost-effective healthcare solutions. The necessity for computationally efficient and highly accurate segmentation algorithms, coupled with the benefits of xLSTM, motivated us to develop the new {\it U-VixLSTM} model for medical image segmentation. Our approach integrates CNNs with ViL for the first time (to our knowledge) for segmentation. The architecture is based on the popular $U-$ shaped framework reported in the literature \cite{rahman2024mist}, \cite{yan2022after}, \cite{dutta2024efficient}, \cite{dutta2023full}. Our model is developed in two different versions: one for handling volumetric inputs and the other for handling 2D medical images. Our research contribution is summarized below. 
\begin{itemize}
    \item  CNNs in the feature extraction path initially capture fine-grained textural information and local patterns corresponding to the target anatomical structures from the input image. The ViL block encodes the global context within the intermediate output volumes, as obtained from the CNN layers. 
    \item The feature reconstruction path upsamples the output from the ViL block to produce the final segmentation output. Skip connections concatenate the feature volume from each level of the feature extraction path to the feature volume of the corresponding level of the feature reconstruction path, to build a robust representation of the target structures.
    \item Experimental results on two publicly available datasets, viz. {\it Synapse} \cite{landman2015miccai}, {\it ISIC} \cite{codella2018skin} and {\it ACDC} \cite{bernard2018deep}, illustrate the effectiveness of our model in terms of performance and utilization of computing resources.
\end{itemize} 
The remaining sections of the paper are structured in the following manner. Section \ref{meth} presents the detailed description of the proposed architecture {\it U-VixLSTM}. The experimental results are illustrated in Section \ref{res}, along with a comparative study with state-of-the-art networks on publicly accessible datasets to validate the efficacy of {\it U-VixLSTM}. Finally, the article is concluded in Section \ref{concl}.

\section{The Architecture} \label{meth}

\begin{figure}[t!]
     \centering
     \includegraphics[width=\linewidth]{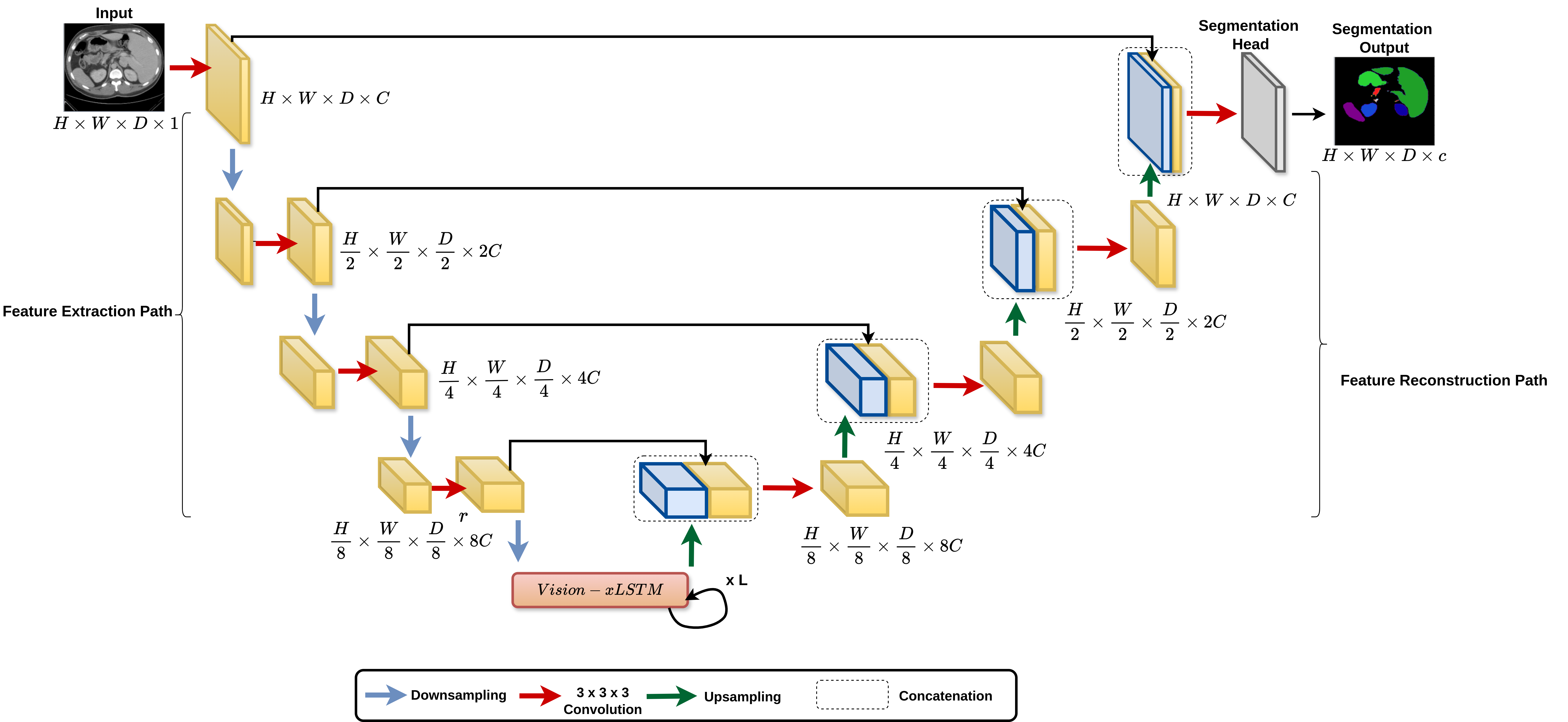}
     \caption{Architectural framework of {\it U-VixLSTM} depicting the input being processed by stacked layers of CNNs and ViL block. The feature representation from the ViL block is upsampled through the feature reconstruction path to obtain the final segmentation output. }
     \label{fig:uvixlstm}
 \end{figure}

The structural framework of our proposed {\it U-VixLSTM} model is depicted in Fig. \ref{fig:uvixlstm}. It follows the classic $U$-shaped framework, characterized by feature extraction and reconstruction pathways. Although the approach is outlined within the framework of 3D volumetric image processing, it can easily be modified for 2D images by reducing spatial dimensions.

The feature extraction arm has multiple layers of CNNs with ViL blocks in the bottleneck. Each ViL block contains the mLSTM layer \cite{beck2024xlstm} to capture long-range dependencies along with temporal awareness. The mLSTM layer employs an exponential gating mechanism to strike a balance between retaining past information and integrating new inputs. The ViL block processes the feature volumes from CNNs to generate an abstract high-level representation of the image. The reconstruction path gradually builds the high-dimensional segmentation output using the contextual representation from the ViL block. The output of each convolution block in the feature extraction path is directed to its corresponding counterpart at the same level of the feature reconstruction path through skip connections. This facilitates the integration of feature maps that originate at various levels of abstraction with the activation maps at the corresponding level of the decoder. It ensures a judicious combination of the finer textural details from earlier convolution levels with the coarser semantic information of the deeper levels, thereby resulting in enhanced context-sensitive predictions. 

\subsection{Feature extraction path}

This path consists of key components, {\it viz.} CNNs for high-level feature learning and ViL for capturing global dependencies.

\subsubsection{High-level features}

The volumetric input image $I \in \mathbb{R}^{H \times W \times D \times 1}$ is passed through a series of convolution layers to hierarchically construct an intermediate abstract and high-level representation of the image denoted by $r \in \mathbb{R}^{\frac{H}{8} \times \frac{W}{8} \times \frac{D}{8} \times 8C}$. Here, $H$, $W$ and $D$ represent the height, width, and depth of the intermediate feature volume, respectively, with $C$ corresponding to the number of channels. 

Next, $r$ is divided into $P \times P \times P$ non-overlapping patches. This is followed by the flattening of these patches into 1D vectors to yield a tokenized representation $\mathbf{t} \in \mathbb{R}^{N \times (P^3\frac{C}{4})}$. Here, $N = (\frac{H}{4} \times \frac{W}{4} \times \frac{D}{4})/P^3$ denotes the number of flattened patches with dimension $P^3 \, \frac{C}{4}$. The flattened patches ($t^1,t^2,\ldots,t^N$) are then projected into a $Z$-dimensional embedding space, with learnable positional embeddings being added to preserve spatial location information. Mathematically, this is expressed as
\begin{equation}
    \mathbf{p} = [t^1\mathbf{K};t^2\mathbf{K};\ldots;t^N\mathbf{K}] + \mathbf{K_{pos}},
\end{equation}
where $\mathbf{K} \in \mathbb{R}^{(P^3\frac{C}{4} \times Z)}$ is the projection matrix and $\mathbf{K_{pos}} \in \mathbb{R}^{N \times Z}$ is the position embedding matrix. $\mathbf{p} \in \mathbb{R}^{N \times Z}$ represents the matrix of flattened patches.
\begin{figure*}[t!]
     \centering
     \includegraphics[height=0.8\textwidth]{figs/mLSTM.png}
     \caption{Architectural framework of the mLSTM layer to compute inter-patch dependencies.}
     \label{fig:mlstm}
 \end{figure*}
\subsubsection{Global dependencies}

The projected patches are processed by the ViL blocks, with the even-numbered blocks handling patch tokens from the top left to the bottom right, and the odd-numbered ones from the bottom right to the top left. Such bidirectional processing enables ViL to capture robust global dependencies in the input. Inspired by the feed-forward network in \cite{vaswani2017attention}, the flattened patches, after normalization, are projected onto an embedding space to increase their dimension by a factor of 2. This step enhances the capability of the model to learn complex and non-linear relationships. These expanded embeddings are divided into two paths, $x_{mlstm} \in \mathbb{R}^{N \times 2Z}$ and $y \in \mathbb{R}^{N \times 2Z}$, as shown in Fig. \ref{fig:mlstm}. $x_{mlstm}$ is further processed by the mLSTM layer of the ViL block.

The mLSTM layer depicted in Fig. \ref{fig:mlstm} is responsible for modeling the inter-patch dependencies. It is an enhanced variant of LSTM that features a matrix memory cell state $\mathbf{C_t}$ rather than a scalar value. At a given time step $t$, $x_{mlstm}$ undergoes a 1D causal convolution with SiLU activation \cite{elfwing2018sigmoid}. This step enriches the patch representation by incorporating information from immediate neighbours. The intermediate result ($\mathbf{X_t} \in \mathbb{R}^{N \times 2Z}$), which is the current input, is then mapped onto query ($q_t$), key ($k_t$) and value ($v_t$) vectors. The generation of query, value and keys is mathematically represented as
\begin{equation}
\mathbf{Q_t} = \mathbf{X_t}\mathbf{W_{Q_t}}^T, \quad \mathbf{K_t} = \mathbf{X_t}\mathbf{W_{K_t}}^T, \quad \mathbf{V_t} = \mathbf{X_t}\mathbf{W_{V_t}}^T,
\end{equation}
where \(\mathbf{Q_t} \in \mathbb{R}^{N \times 2Z}\), \(\mathbf{K_t} \in \mathbb{R}^{N \times 2Z}\),\(\mathbf{V_t} \in \mathbb{R}^{N \times 2Z}\) are the query, key and value matrices, respectively. Here, \(\mathbf{W_{Q_t}} \in \mathbb{R}^{2Z \times 2Z}\), \(\mathbf{W_{K_t}} \in \mathbb{R}^{2Z \times 2Z}\), \(\mathbf{W_{V_t}} \in \mathbb{R}^{2Z \times 2Z}\) are learnable weight matrices to generate the query, key, and value vectors.

The input and forget gate pre-activations, $\mathbf{i_t} \in \mathbb{R}^{N \times 2Z}$ and $\mathbf{f_t} \in \mathbb{R}^{N \times 2Z}$ are simultaneously calculated from $\mathbf{X_t}$ as 
\begin{equation}
    \mathbf{i_t} = \exp((\mathbf{W^I})^T \mathbf{X_t} + \mathbf{B}),
\end{equation}
\begin{equation}
    \mathbf{f_t} = \exp((\mathbf{W^F})^T \mathbf{X_t} + \mathbf{B}).
\end{equation}
Here, $\mathbf{B} \in \mathbb{R}^{N \times 2Z}$ is the bias matrix. $\mathbf{W^I},\mathbf{W^F} \in \mathbb{R}^{2Z \times 2Z}$ are the projection matrices for $\mathbf{i_t}$ and $\mathbf{f_t}$ respectively. $\exp$ represents the exponentiation operation introduced in xLSTM to overcome the vanishing gradient issue. This enables the model to learn complex relationships among long patch sequences without losing relevant context. The input gate decides which information to store in the current matrix memory cell state $\mathbf{C_t}$. The forget gate is responsible for discarding information from the previous cell state at timestamp $t-1$. The cell state $\mathbf{C_t}$ is updated as follows:\\
\begin{equation}
    \mathbf{C_t} = \mathbf{i_t}.\mathbf{V_t}\mathbf{K_t}^T+\mathbf{f_t}.\mathbf{C_{t-1}}  
\end{equation}
$\mathbf{V_t}\mathbf{K_t}^T$ is the key-value pair representing the matrix of current input information to be stored at time-step $t$.
An additional normalizer state $\eta_t$ is introduced to tackle the exploding cell state values due to the exponentiation operation. The normalizer state pre-activation is computed as follows:
\begin{equation}
    \eta_t = \mathbf{f_t}.\eta_{t-1} + \mathbf{i_t}.\mathbf{K_t}
\end{equation}
The normalizer states, until time-step $t-1$, are scaled by $\mathbf{f_t}$. This determines the influence of the previous normalizer states on the current normalizer state. The contribution of the current state input $\mathbf{K_t}$ is controlled by $\mathbf{i_t}$.

Subsequently, the hidden state $h_t$ representing the output of the current time-step $t$ is calculated as follows:
\begin{equation}
    h_t = o_t[\mathbf{C_t/\eta_t}]
\end{equation}
 The memory cell state $\mathbf{C_t}$ is normalized by $\eta_t$ to prevent the exponential growth of cell state values, thereby ensuring numerical stability. $o_t = \sigma (\mathbf{W^o{^T} X_t} + \mathbf{B})$ represents the output gate activation, which selectively filters the information from normalized $\mathbf{C_t}$ to be written to $h_t$. The sigmoid activation function, denoted by $\sigma$, squashes $o_t$ in the range $[0,1]$. Multiplication ($\odot$) of $h_t$ with $y$, as shown in Fig. \ref{fig:mlstm}, enhances the patch embedding matrix $y$ with the overall contextual information encoded by $h_t$.

 ViL processes one patch at a time due to the recurrent structure instead of attending to all the previous patches like ViTs. Thereby, the number of computations remains constant for each patch, which leads to an overall computational complexity of $\mathcal{O}(n)$, where $n$ is the total number of patches. ViL retains a constant-size internal memory throughout, regardless of the total number of patches present, instead of storing the entire set of key-value pairs corresponding to different patches.

\subsection{Feature reconstruction path}

A trilinear upsampling operation \( \tau \) is employed, at every level $l$, to increase the spatial dimension of the feature maps obtained from the previous level \( l+1 \). This helps align the spatial dimensions of the feature maps with those received from the corresponding level $l$ of the feature extraction path. The upsampled feature map at level \( l \) is expressed as
\begin{equation}
\mathbf{U}_l = \tau(\mathbf{F}_{l+1}),
\end{equation}
where  \( \mathbf{F}_{l+1} \) is the output volume from level \( l+1 \).

The upsampled feature map \( \mathbf{U}_l \) is then concatenated with the feature volume from the corresponding level \( l \) of the feature extraction path, denoted as \( \mathbf{E}_l \). The concatenated feature volume at level \( l \) is given by
\begin{equation}
\mathbf{C}_l = \text{Concat}(\mathbf{U}_l, \mathbf{E}_l).
\end{equation}
It is then convolved to yield the output volume \( \mathbf{R}_l \) at level \( l \) of the feature reconstruction path, as 
\begin{equation}
\mathbf{R}_l = \text{Conv}_{3 \times 3 \times 3}(\mathbf{C}_l).
\end{equation}

The training procedure for {\it U-VixLSTM} is described in Algorithm \ref{alg:u_vixlstm}.

\begin{algorithm}[h!]
\footnotesize
\caption{Training Process for {\it U-VixLSTM}}
\label{alg:u_vixlstm}
\textbf{Input:} Training dataset $\Delta=\{(I_j,O_j)\}_{j=1}^m$, Patch size $P$, Embedding dimension $Z$, Number of ViL blocks $L$, Projection matrix $K$, Positional embedding $K_{pos}$, learning rate $\epsilon$ and number of iterations $\iota$ \\
\textbf{Output:} Segmentation output $\{\hat{O_j}\}_{i=1}^m$

\begin{algorithmic}[1]
    \State Initialize weights $\Omega$
    \While{iterations $\leq$ $\iota$} 
        \State /* Feature Extraction Path */
        \For{$l=1$ to $x$} /* Iterating through different levels of feature extraction path */
            \State $E_l = Conv(E_{l-1})$ /* Apply convolution */
            \State $E_l = ReLU (E_l)$
            \State $E_l = Downsample (E_l)$
        \EndFor
        \State $r \gets E_x$ /* Intermediate feature representation $r$ from a series of convolution blocks at level $x$ */ \\
        /* Divide $r$ into $N = \frac{HWD}{P^3}$ patches of size $P^3$ */
        \State $conv \gets$ Conv3D (filters = $\frac{HWD}{P^3}$, kernel size = $(P,P,P))$
        \State $ patches = conv(r)$
        \State Flatten patches into tokens $t$
        \State /* Project tokens to embedding space $Z$ */
        \State $p \gets []$
        \For{$i=1$ to $N$}
            \State p.append($t_i * K$)
        \EndFor
        \State $p \gets p + K_{pos}$
        \For{$i=1$ to $L$} /* Apply ViL block */
            \State Compute memory cell state $\mathbf{C_t}$ and gate activations with normalizer state
            \State Update hidden state $h_t$ 
        \EndFor
        \State /* Feature Reconstruction Path */
        \For{$l=x-1$ to $1$}
            \State $U_l \gets \tau(F_{l+1})$
            \State $C_l \gets Concat(U_l, E_l)$
            \State $R_l \gets Conv(C_l)$
        \EndFor

        \State $\hat{O} \gets Conv_{1\times1\times}(R_1)$
        \State Compute loss $\mathbf{L}(\{O,\hat{O}\})$
        \State $\Omega' \gets \Omega-\epsilon\frac{\partial\mathbf{L}}{\partial \Omega}$ /*Weight updation*/
     \EndWhile   

\end{algorithmic}
\end{algorithm}

\section{Experimental Details} \label{res}

This section describes the loss functions and performance metrics used for the evaluation, details of the datasets used and the experimental results. The PyTorch and MONAI frameworks were used to implement and train {\it U-VixLSTM} in Python 3.9 on an NVIDIA RTX A6000 GPU with 48GB of RAM. AdamW Optimizer was used, with a learning rate of 1e-4 and a weight decay of 1e-5. 

\subsection{Loss function and metrics}

A combination of Dice $\mathcal{L}_{dice} = \gamma - \sum_{\gamma=1}^c\left(\frac{2 \sum_{i=1}^T \hat{z}_{\gamma,i}z_{\gamma,i} + \mu}{\sum_{i=1}^T \hat{z}_{\gamma,i}+z_{\gamma,i} + \mu}\right)$ \cite{milletari2016v} and Categorical Cross Entropy \cite{taghanaki2019combo} $\mathcal{L}_{cce} = -\frac{1}{T}\sum_{i=1}^T\sum_{\gamma=1}^c z_{\gamma,i}\log{(\hat{z}_{\gamma,i})}$ was employed to train the model. Here, $c$ represents the total number of classes to be predicted,  $\hat{z}_{\gamma, i}$ and $z_{\gamma,i }$ correspond to the predicted and ground truth values (respectively) for the $i^{th}$ voxel for class $\gamma$. $T$ denotes the total number of voxels in the input and $\mu$ is the additive smoothing parameter used to overcome division-by-zero errors.
The composite loss function $\mathcal{L}$ is expressed as
\begin{equation}
     \mathcal{L}(\{\hat{\rho},\rho\};\Gamma)=\mathcal{L}_{dice}(\{\hat{\rho},\rho\},\Gamma)+\mathcal{L}_{cce}(\{\hat{\rho},\rho\},\Gamma),
\end{equation}
with $\Gamma$ denoting the model parameters, $\hat{\rho}$ and $\rho$ symbolizing  predicted segmentation map and ground truth, respectively.

The Dice Score Coefficient ($DSC$), Intersection-over-Union ($IoU$), and 95\% Hausdorff Distance ($HD95$) were the performance metrics used to evaluate the segmentation output. 

\begin{figure*}[t]
    \centering
    \includegraphics[width=\linewidth]{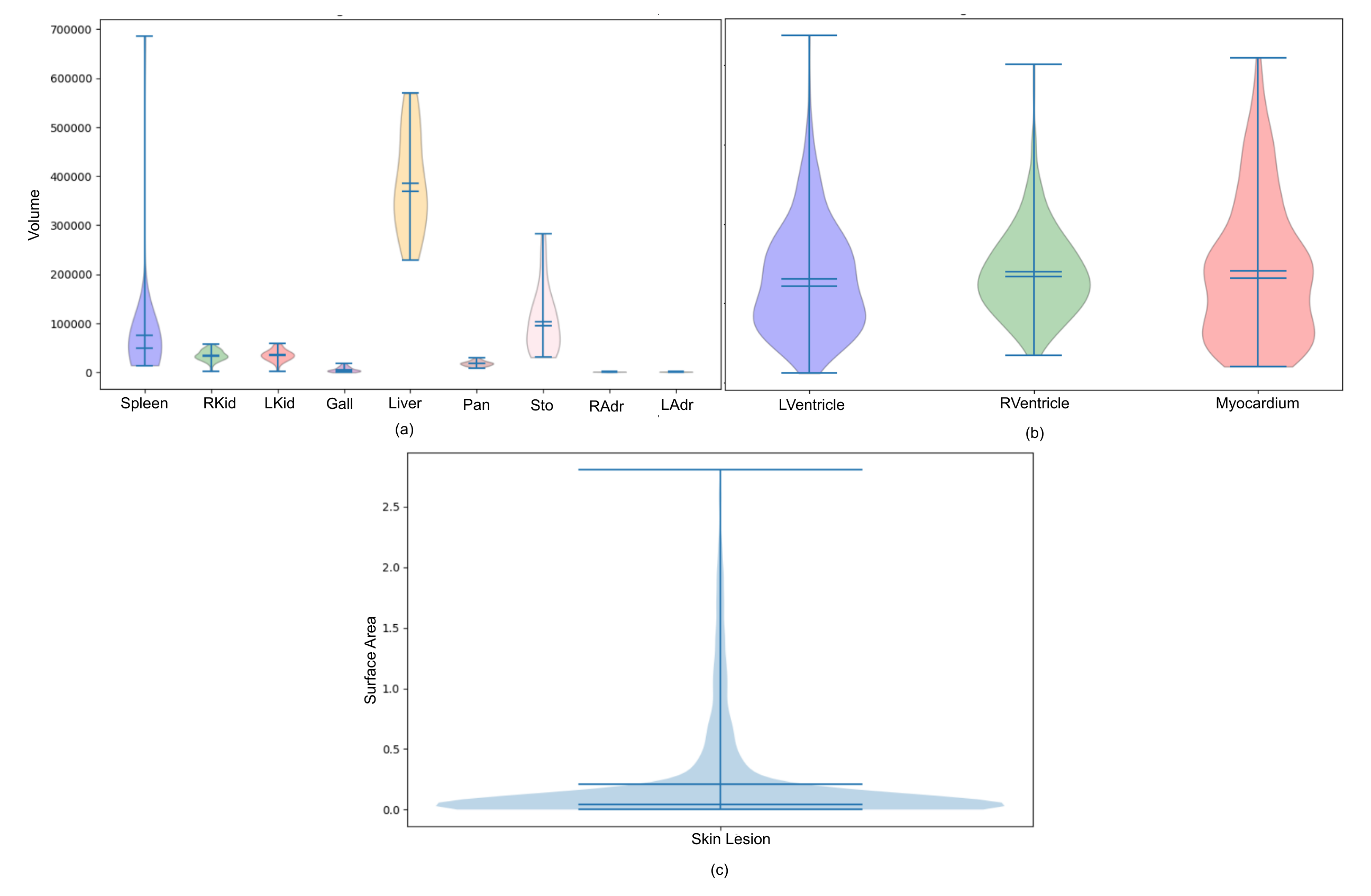}
    \caption{Distribution of (a) various abdominal organs, (b) cardiac anatomical structures, and (c) skin lesion surface area, among different patients in the {\it Synapse}, {\it ACDC} and {\it ISIC} datasets, respectively.}
    \label{fig:dist}
\end{figure*}

\subsection{Datasets}

The data set {\it Synapse} comprises 30 CT volumes with sizes ranging from 512 $\times$ 512 $\times$ 85 to 512 $\times$ 512 $\times$ 198.  CT volumes are manually annotated by experts in the field to highlight the different abdominal organs. A model is trained to segment nine distinct organs of the abdominal cavity, {\it viz.} spleen, left kidney, right kidney, liver, gall bladder, pancreas, stomach, right adrenal gland and left adrenal gland. The spleen, liver, and stomach are classified as larger organs, while the kidneys, gall bladder, pancreas, and adrenal glands are smaller in size.
The CT volumes were divided into training and test sets, with 80\% used for training and the remaining 20\% used for testing to evaluate the generalization performance of the model. The voxel intensities were restricted between -170 HU and 250 HU, followed by normalization of the intensity to the range of [0,1]. Data augmentation was used to expand the size of the training dataset. 

The {\it ISIC} dataset contains dermoscopic images with pixel values $\in [0,255]$. These images have been curated by the International Skin Imaging Collaboration (ISIC) for the study of skin cancer. The training dataset consists of 2000 images with corresponding ground truth masks prepared by domain experts. The test set consists of 600 images. The value of pixel 0 in the ground truth mask corresponds to the background region, whereas the value of pixel 255 corresponds to the lesion region. The input images and their corresponding masks were normalized to the pixel values $\in [0,1]$. The training data was augmented by rotation and random cropping transformation. 

The dataset {\it ACDC} consists of 150 volumes of chest magnetic resonance images for automated cardiac diagnosis. Magnetic resonance volumes were obtained from the University Hospital of Dijon, France. The experts prepared the corresponding ground truth volumes, which have segmentations for the Right Ventricle (RV), Left Ventricle (LV), and Myocardium. Volumes were pre-processed analogously to that of the {\it Synapse} dataset.

Fig. \ref{fig:dist} illustrates the distribution of abdominal and cardiac organ volumes, as well as the surface area of skin lesions, between patients in the three datasets. The plots demonstrate substantial evidence of inter-patient variability across the datasets, indicating significant differences in the sizes of various organs or lesion areas among patients. Quite a substantial difference in morphology is evident in the spleen, liver, and stomach, as shown in Fig. \ref{fig:dist}(a), in the cardiac organs of Fig. \ref{fig:dist}(b), and in the total surface area of the lesion illustrated in Fig. \ref{fig:dist}(c), as attributed to long tails and broader distributions. The median values indicate significant heterogeneity in the average volumes of the various abdominal organs. The liver exhibits the largest average volume, followed by the stomach and spleen. The adrenal gland exhibits the lowest average volume among the glands. The diverse sizes of target structures hinder the ability of the model to generalize across datasets. Fig. \ref{fig:dist}(c) illustrates a right-skewed distribution of the surface area of the lesions. This suggests that patients generally have smaller lesion areas compared to larger ones. This may indicate a class imbalance within the dataset.

The volumetric samples of {\it Synapse} and {\it ACDC} are processed by the 3D version of {\it U-VixLSTM} while the 2D version is trained using the samples from {\it ISIC} dataset.

\subsection{Results and Discussion}

\begin{table}[t]
\centering
\caption{Comparison of different variants of {\it U-VixLSTM} with increasing number of ViL blocks ($\times L$) and convolution layers, on {\it Synapse} data. Best results and the selected model configuration are highlighted in \textbf{bold}.}
\label{table:abl}
\setlength{\tabcolsep}{3pt}
\resizebox{0.5\linewidth}{!}{\begin{tabular}{c|c|c|c|c}
\hline
\textbf{Ablations}                               & \multicolumn{1}{l|}{\begin{tabular}{c}\textbf{Model} \\ \textbf{Variants}
\end{tabular}} & \multicolumn{1}{l|}{\textbf{mDSC}} & \multicolumn{1}{l|}{\textbf{mIoU}} & \multicolumn{1}{l}{\textbf{mHD95}} \\ \hline
\multirow{5}{*}{ \begin{tabular}{c}
\# ViL \\ blocks
\end{tabular}} & x 3                        & 0.8118          & 0.7156          & 5.56                       \\
                                        & \textbf{x 6}                        & \textbf{0.8318}          & \textbf{0.7323}          & 4.80 \\  
                                        & x 12                                & 0.8289                   & 0.7286                   & 8.57                      \\
                                        & x 18                                & 0.8201                   & 0.7189          & \textbf{4.34}             \\
                                        & x 24                                & 0.8299                   & 0.7280                    & 19.05                     \\ \hline
\multirow{3}{*}{\begin{tabular}{c}
\# Convolution \\Layers
\end{tabular} }  & 3                                   & 0.8143                   & 0.7114                   & 5.48                      \\
                                        & \textbf{4}                          & \textbf{0.8318}          & \textbf{0.7323}         & \textbf{4.80}                       \\
                                        & 5                                  & 0.8314                   & 0.7315                   & 4.85 \\ \hline          
\end{tabular}}
\end{table}

Table \ref{table:abl} presents an ablation study on {\it Synapse} data, demonstrating the impact of different numbers ($L$) of ViL blocks and a varying number of convolution layers along the feature extraction path.  The average scores of $DSC$, $HD95$, and $IoU$ are reported for the different organs. The ablation study reflects the trade-off between the complexity of the model and its performance. The best $DSC$ and $IoU$ values were observed for 6 ViL blocks. The performance degrades when increasing the number of ViL blocks beyond 6, which indicates overfitting. The model appears to memorize the training data instead of learning generalizable features, which occurs as the number of parameters increases due to the addition of more ViL blocks. The 6-block configuration appears to be sufficient on this dataset to capture the global context without the risk of overfitting. Similarly, increasing the number of convolution layers from 3 to 4 shows a significant gain in metric scores. This indicates the importance of hierarchical depth in extracting rich representations. However, the performance saturates with a further increase in the number of layers, which indicates minimal benefits with an increase in the computational load. The two-part ablation study empirically justifies the design choice of {\it U-VixLSTM} to extract robust spatial features hierarchically, followed by modeling global context. This approach efficiently mitigates the risk of overfitting while ensuring computational feasibility.  


\begin{table*}[t]
\caption{Comparison with state-of-the-art models on multi-organ segmentation (Synapse) dataset, with best results marked in \textbf{bold}.}
\label{table:synapse}
\centering
\renewcommand\arraystretch{1.2}
\resizebox{\linewidth}{!}{\begin{tabular}{c|ccccccccc|c|c|c}
\toprule
   & \multicolumn{10}{c|}{\textbf{DSC}}    & \textbf{IoU}             & \textbf{HD95}                          \\ \cline{2-11}

\multirow{-2}{*}{\textbf{Model}}                                    & \textbf{Spleen}               & \textbf{Right kidney}         & \textbf{Left kidney}          & \textbf{Gall Bladder}         & \textbf{Liver}                & \textbf{Pancreas}                      & \textbf{Stomach}                       & \textbf{Right Adrenal}                 & \textbf{Left Adrenal}                  & \textbf{Mean}            & \textbf{Mean}            & \textbf{Mean}                          \\ \hline
$U$-Net                                                      & 0.9112                        & 0.9007                        & 0.9181                        & 0.5645                        & 0.9572                        & 0.6967                        & 0.7800                          & 0.6160                         & 0.5872                        & 0.7702          & 0.6667          & 38.83                         \\
$V$-Net                                                     & 0.8874                        & 0.9251                        & 0.9244                        & 0.5858                        & 0.9487                        & 0.7511                        & 0.7953                        & 0.6004                        & 0.4497                        & 0.7631          & 0.6537          & 25.29                         \\
$U$-Net++                                                    & 0.9118                        & 0.9196                        & 0.8905                        & 0.6921                        & 0.9524                        & 0.7536                        & 0.7869                        & 0.5245                        & 0.1032                        & 0.7261          & 0.6254          & 55.27                         \\
Attention $U$-Net                                            & 0.9109                        & 0.8770                         & 0.8720                         & 0.5835                        & 0.9585                        & 0.5566                        & 0.7846                        & 0.5991                        & 0.5398                        & 0.7424          & 0.6262          & 51.61                         \\
$U$-Net + EfficientNet-b0   &  0.8541 &  0.8919 &  0.8804 & 0.6370  &  0.9077 &  0.5018 &  0.7097 & 0.6048 & 0.5195 & 0.7230           & 0.6056          & 67.47                         \\
$U$-Net + EfficientNet-b1  &  0.7414 &  0.8059 &  0.8256 & 0.5500   & 0.9019 &  0.5416 &  0.6160  &  0.5250  &  0.5035 & 0.6679          & 0.5539          & 75.85                         \\
$U$-Net 3+                                          &  0.8736 &  0.9229 & 0.8584 & 0.6659 &  0.9383 & 0.6823 & 0.7259                        & 0.5657                        & 0.4764                        & 0.7455          & 0.6254          & 54.43                         \\ 
Swin UNETR                                                  & 0.9482                        & 0.9300                          & 0.9245                        & 0.7617                        & 0.9622                        & 0.8046                        & 0.8059                        & \textbf{0.6890}                & 0.6159                        & 0.8269          & 0.7261          & 13.99                         \\
UNETR                                                    & 0.8951                        & 0.9055                        & 0.8950                         & 0.5274                        & 0.9461                        & 0.6668                        & 0.7847                        & 0.5268                        & 0.4978                        & 0.7384          & 0.6284          & 26.94                         \\
TransUNet                                                 & 0.8520                         & 0.8828                        & 0.7850                         & 0.7608                        & 0.5870                         & 0.7218                        & \textbf{0.8773}                      & 0.5876                        & 0.4672                        & 0.7246          & 0.6679          & 36.81                         \\
TransAttUNet                                            & 0.9045                        & 0.8761                        & 0.9160                         & 0.4958                        & 0.9408                        & 0.5997                        & 0.6682                        & 0.6559                        & 0.5948                        & 0.7391          & 0.6580           & 15.82                         \\
Swin-UMamba                                              & 0.8778                        & 0.9037                        & 0.9112                         & 0.4206                       & 0.9161                        & 0.6238                        & 0.6527                       & 0.5890                        & 0.5617                        & 0.7174          & 0.6334           & 20.10                         \\
UNETR++       & 0.8061                        & 0.8050                        & 0.8201                        & 0.5208                       & 0.8868                        & 0.5249                        & 0.5576                       & 0.5890                        & 0.4577                        & 0.6631        & 0.5340           & 8.61                         \\
DS-TransUNet                                              & 0.4965                        & 0.6113                        & 0.5252                        & 0.5397                        & 0.6790                         & 0.5295                        & 0.5857                        & 0.5745                        & 0.5266                        & 0.5631          & 0.4644          & 14.79 \\
\midrule
\textit{\textbf{U-VixLSTM}}                           & \textbf{0.9500}                 & \textbf{0.9371}               & \textbf{0.9366}               & \textbf{0.8104}                        & \textbf{0.9635}               & \textbf{0.7878}               & \underline{0.8304}               & \underline{0.6709}                        & \textbf{0.6458}               & \textbf{0.8318} & \textbf{0.7286} & \textbf{4.80}  \\
\bottomrule                      
\end{tabular}}
\end{table*}

The performance of our {\it U-VixLSTM} was next compared with that of the state-of-the-art (SOTA) algorithms in the literature in terms of metrics $DSC$, $IoU$, $HD95$ on {\it Synapse}, {\it ISIC} and {\it ACDC} datasets. Table \ref{table:synapse} presents a comprehensive breakdown of performance for different organs (in {\it Synapse} data) in the context of {\it DSC}. The average results for $HD95$ and {\it IoU} are provided for the nine abdominal organs. {\it U-VixLSTM} is found to outperform other SOTA,  with respect to the mean {\it DSC}, {\it IoU} and $HD95$, with scores of 83.18\%, 72.86\%, and 4.8, respectively.  Our proposed {\it U-VixLSTM} demonstrated superior performance in generating the highest $DSC$ values for the segmentation of larger organs (such as the spleen, liver) and smaller organs (such as the kidneys, pancreas, gall bladder and left adrenal gland). Performance remained stable, despite the reduction in organ size, as indicated by the consistently high $DSC$ scores observed in smaller and larger organs. This illustrates the robustness of our model in utilizing acquired knowledge on anatomical structures with varying shapes and sizes. It is found to precisely identify and define the target areas possessing unique structures.

\begin{figure*}[t!]
    \centering
    \includegraphics[width=\linewidth]{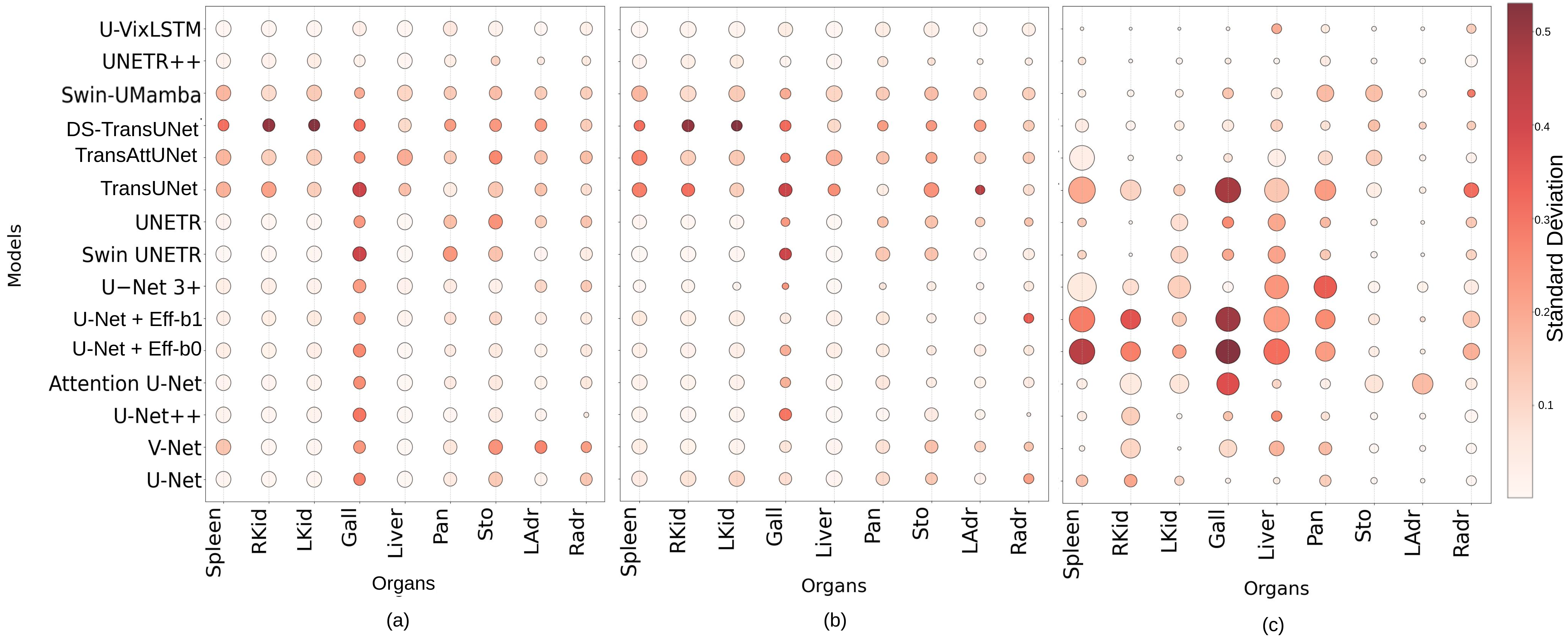}
    \caption{Dot plot of (a) {\it DSC}, (b) {\it IoU}, and (c) $HD95$ metrics for evaluating the performance of {\it U-VixLSTM} against other baselines on {\it Synapse} dataset. The radius of the circles indicates the metric values, while the color intensity signifies the standard deviation.}
    \label{fig:dot_btcv}
\end{figure*}

\begin{figure*}[h!]
    \centering
    \includegraphics[height=\linewidth]{figs/qualitative_res_BTCV_9_classes_UVixLSTM.png}
    \caption{Comparative performance of {\it U-VixLSTM} with other baseline architectures, on the {\it Synapse} dataset, through sample segmentation maps. The first row in each block represents a sample CT slice. The second row in each block provides zoomed-in boxes to provide a magnified view of specific regions. (a) Input CT image, (b) corresponding ground truth, with the respective output from (c) {\it U-VixLSTM}, (d) Swin UNETR, (e) UNETR, (f) {\it V-}Net, and (g) TransAttUNet.}
    \label{fig:btcv_9}
\end{figure*} 

Fig. \ref{fig:dot_btcv} visualizes performance consistency across different metrics for {\it U-VixLSTM} and baselines concerning different abdominal organs. The visualization represents the metric values (denoted by the dot radius) and consistency (represented by the color intensity of the dots). {\it U-VixLSTM} exhibits consistent superiority across various organs, as evidenced by the larger radii and uniformly pale color intensities of the dots in Fig. \ref{fig:dot_btcv}(a) and (b). This signifies higher accuracy with low standard deviation. Therefore, the high accuracy of the proposed model is both consistent and reliable. Conversely, Swin-UMamba, DS-TransUNet, TransAttUNet and TransUNet, having darker and smaller dots, signifies high variability and lesser accuracy than {\it U-VixLSTM}. Fig. \ref{fig:dot_btcv}(c) represents the $HD95$ metric values across all the organs for different models. A smaller dot radius means higher accuracy in delineating the boundaries for this metric. {\it U-VixLSTM} consistently demonstrates smallest dot radii with pale color intensities for nearly all the organs. This indicates the superior boundary delineating capabilities compared to baselines, like {\it U}-Net 3+, {\it U}-Net with EfficientNet backbone, Attention {\it U-}Net and TransUNet, having significantly larger and darker dots. While TransUNet attains the highest mean {\it DSC} value for the stomach, it exhibits a considerably greater standard deviation and higher $HD95$ values in comparison to {\it U-VixLSTM}. This indicates that while TransUNet effectively segments the majority of the stomach region, it encounters difficulties in accurately delineating the organ boundary and sustaining consistent performance.

Fig. \ref{fig:btcv_9} shows the sample segmentation output of {\it U-VixLSTM} and other architectures, such as Swin UNETR, UNETR, $V$-Net, and TransAttUNet, on the {\it Synapse} dataset. It is evident that the {\it U-VixLSTM} model demonstrates a significantly higher level of similarity to the ground truth, compared to the maps produced by these other baseline models. The results of the Swin UNETR and UNETR models show $FP$ regions corresponding to the stomach, in the second row of Figs. \ref{fig:btcv_9}(d)-(e). The results of $V$-Net demonstrate a lack of precision in identifying the anatomical regions associated with the adrenal gland, kidneys, pancreas and stomach. The segmentation maps obtained from TransAttUNet exhibit their limited capacity to learn complex representations corresponding to multiple organs, with varying shapes and sizes.

The hybrid architecture of {\it U-VixLSTM} shows effectiveness in segmenting abdominal organs with diverse shapes and sizes. The CNN layers effectively capture intricate details in high-resolution maps, which is essential for accurately delineating small-sized organs. The ViL block strongly models the global context, making it well-suited for the segmentation of larger organs.

\begin{figure*}[t]
    \centering
    \includegraphics[width=\linewidth]{ISIC_qualitative_selected.png}
    \caption{Comparative performance of {\it U-VixLSTM} and other baseline architectures, on the {\it ISIC} dataset, through sample segmentation maps. (a) Input dermoscopic image, (b) corresponding ground truth, with the respective output from (c) {\it U-VixLSTM}, (d) Swin UNETR, (e) UNETR, (f) {\it U}-Net 3+, (g) TransAttUNet, and (h) DS-TransUNet.}
    \label{fig:isic}
\end{figure*}

\begin{table}[t!]
\caption{Comparison with state-of-the-art models on the {\it ISIC} dataset, with best results marked in \textbf{bold}. }
\label{table:isic}
\centering
\resizebox{0.4\linewidth}{!}{\begin{tabular}{c|c|c|c}
\hline
\textbf{Model}     & \textbf{DSC}     & \textbf{IOU}    & \textbf{HD95}   \\
\hline
$U-$Net       & 0.8065      & 0.7197     & 74.72        \\
Attention $U-$Net    & 0.7798     & 0.6845    & 87.13      \\

$U-$Net + & & &\\
EfficientNet-b0                         & \multicolumn{1}{c|}{ 0.8222} & 0.7315                                            & \multicolumn{1}{c}{{12.32}} \\
$U-$Net + & & &\\
EfficientNet-b1  & \multicolumn{1}{c|}{ 0.8120}  & \multicolumn{1}{c|}{ 0.7176} & \multicolumn{1}{c}{ 13.44} \\
$U-$Net ++ & 0.8074 & 0.7057 & 13.37\\
LB-UNet & 0.8092 & 0.7164 & 57.05 \\ 
$U-$Net 3+ &  0.7732 & 0.6772 & 14.77 \\
Swin UNETR                                                 & 0.8187                                             & 0.7289                                             & 61.06                                            \\
TransUNet                                                  & 0.7012                                             & 0.6086                                             & 90.32                                            \\
UNETR                                                      & \multicolumn{1}{c|}{0.7601} & \multicolumn{1}{c|}{0.6495} & \multicolumn{1}{c}{19.04} \\
TransAttUNet                                               & 0.7580                                              & 0.6610                                              & 82.22\\
DS-TransUNet & 0.8212 & 0.7304 & 12.50\\
Swin-UMamba & 0.8237 & 0.7330 & 56.41 \\
$\text{Swin-UMamba}^\dagger$ & 0.8264 & 0.7334 & 55.05 \\
\midrule
\textbf{\textit{U-VixLSTM}}                                          & \textbf{0.8500}                                       & \textbf{0.7611}                                    & \textbf{11.31}          \\
\hline
\end{tabular}}
\end{table}

Fig. \ref{fig:isic} represents the sample segmentation output, as predicted by {\it U-VixLSTM} and other baseline models on the {\it ISIC} dataset. The prediction made by our {\it U-VixLSTM} exhibits higher accuracy and precision in relation to the ground truth masks, compared to the maps generated by the other baseline models. The sample outputs from the other state-of-the-art architectures mostly contain undersegmented or oversegmented regions. For example, along rows 2 and 3, the baseline models exhibit a significant amount of $FN$ and $FP$ regions; our {\it U-VixLSTM}, on the other hand,  has a visually higher proportion of $TP$ regions and fewer $FP$ and $FN$ pixels. This indicates that these other models had difficulty accurately identifying the boundaries of the target region(s). The quantitative results, presented in Table \ref{table:isic}, demonstrate the highest values of $DSC$ and $IoU$ along with the lowest $HD95$ metric score for our {\it U-VixLSTM}. This corroborates the qualitative observation of the enhanced segmentation accuracy of our proposed model. 

\begin{figure*}[h!]
    \centering
    \includegraphics[width=\linewidth]{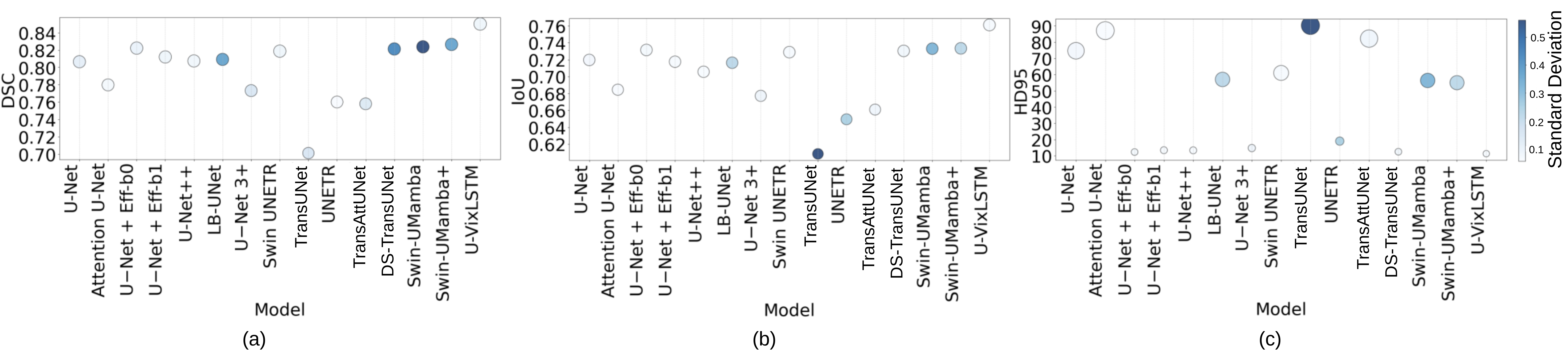}
    \caption{Dot plot of (a) {\it DSC}, (b) {\it IoU}, and (c) $HD95$ metrics for evaluating the performance of {\it U-VixLSTM} against other baselines on {\it ISIC} data. The radius of the circles indicates the metric values, while the color intensity signifies the standard deviation.}
    \label{fig:dot_isic}
\end{figure*}

Fig. \ref{fig:dot_isic} provides a detailed analysis of {\it U-VixLSTM} against the baselines across all the metrics. The pale blue hue of the dot corresponding to {\it U-VixLSTM} across all the subplots signifies a minimal standard deviation. This demonstrates the superior performance of the model across the diverse dermoscopic images. Precisely delineating the boundaries of the affected region is a significant challenge in skin lesion analysis. Fig. \ref{fig:dot_isic}(c) demonstrates the efficacy of {\it U-VixLSTM} in addressing this challenge, as indicated by the smallest dot representing the minimum $HD95$ value of 11.31. On the contrary, Transformer or Mamba-based models exhibit a larger radius (suboptimal boundary delineation), with darker blue hues signifying discrepancies in boundary prediction.

\begin{table}[t]
\caption{Comparison with state-of-the-art models on the {\it ACDC} dataset, with best results marked in \textbf{bold}. }
\label{table:acdc}
\centering
\resizebox{0.8\linewidth}{!}{
\begin{tabular}{c|ccc|c|c|c}
\hline
                                                           & \multicolumn{4}{c|}{\textbf{DSC}}                                                                                   &                        &                       \\ \cline{2-5}
\multirow{-2}{*}{\textbf{Model}}                                    & \multicolumn{1}{l}{\textbf{LVentricle}} & \multicolumn{1}{l}{\textbf{RVentricle}} & \multicolumn{1}{l|}{\textbf{Myocardium}} & \textbf{Mean}   & \multirow{-2}{*}{\textbf{mIOU}} & \multirow{-2}{*}{\textbf{mHD95}} \\ \hline
$U$-Net                                                       & 0.7869                         & 0.7956                         & 0.8910                          & 0.8245 & 0.7130                  & 7.29                  \\
$V$-Net                                                      & 0.7273                         & 0.6969                         & 0.8108                         & 0.7450  & 0.6145                 & 6.34                  \\
$U$-Net++                                                     & 0.6155                         & 0.6828                         & 0.8302                         & 0.7095 & 0.5693                 & 54.35                 \\
Attention $U$-Net                                            & 0.7087                         & 0.7637                         & 0.8689                         & 0.7804  & 0.6568                 & 8.32                  \\
$U$-Net + EfficientNet-b0 backbone                         & 0.6236                         & 0.5314                         & 0.7069                         & 0.6206 & 0.5219                 & 104.65                \\
$U$-Net + EfficientNet-b1 backbone & 0.6543                         & 0.5217                         & 0.7299                         & 0.6353 & 0.5026                 & 98.27                 \\
$U$-Net 3+                                                    & 0.6903                         & 0.6412                         & 0.7355                         & 0.6890  & 0.5463                 & 149.64                \\ 
Swin UNETR                             & 0.8059                         & 0.7741                         & 0.8838                         & 0.8213 & 0.7076                 & 6.02                  \\
UNETR                      & 0.7052                         & 0.6845                         & 0.8316                         & 0.7404 & 0.6049                 & 9.84                  \\
TransUNet                                                 & 0.6398                         & 0.7692                         & 0.8714                         & 0.7601 & 0.6581                 & 6.93                  \\
TransAttUNet                                               & 0.6338                         & 0.7420                          & 0.8819                         & 0.7526 & 0.6432                 & 95.35                 \\
Swin-UMamba                                                & 0.6640                          & 0.7907                         & 0.8699                         & 0.7749 & 0.6720                  & 5.45                 \\
DS-TransUNet                                              & 0.6396                         & 0.7688                         & 0.8654                         & 0.7579 & 0.6540                  & 5.32                  \\
UNETR++                                                    & 0.7026                         & 0.6977                         & 0.8264                         & 0.7422 & 0.6121                 & 6.59                  \\ \hline
\textit{\textbf{U-VixLSTM}}                                & \textbf{0.8680}                          & \textbf{0.8345}                         & \textbf{0.9104}                         & \textbf{0.8710}  & \textbf{0.7770}                  & \textbf{5.07}         \\ \hline        
\end{tabular}}
\end{table}
\begin{figure*}[!h]
    \centering
    \includegraphics[width=\linewidth]{figs/UVixLSTM_dotplot_acdc.png}
    \caption{Dot plot of (a) {\it DSC}, (b) {\it IoU}, and (c) $HD95$ metrics for evaluating the performance of {\it UVixLSTM} against other baselines on {\it ACDC} data. The radius of the circles indicates the metric values, while the color intensity signifies the standard deviation.}
    \label{fig:dot_acdc}
\end{figure*}
\begin{figure*}[h!]
    \centering
    \includegraphics[width=0.5\linewidth]{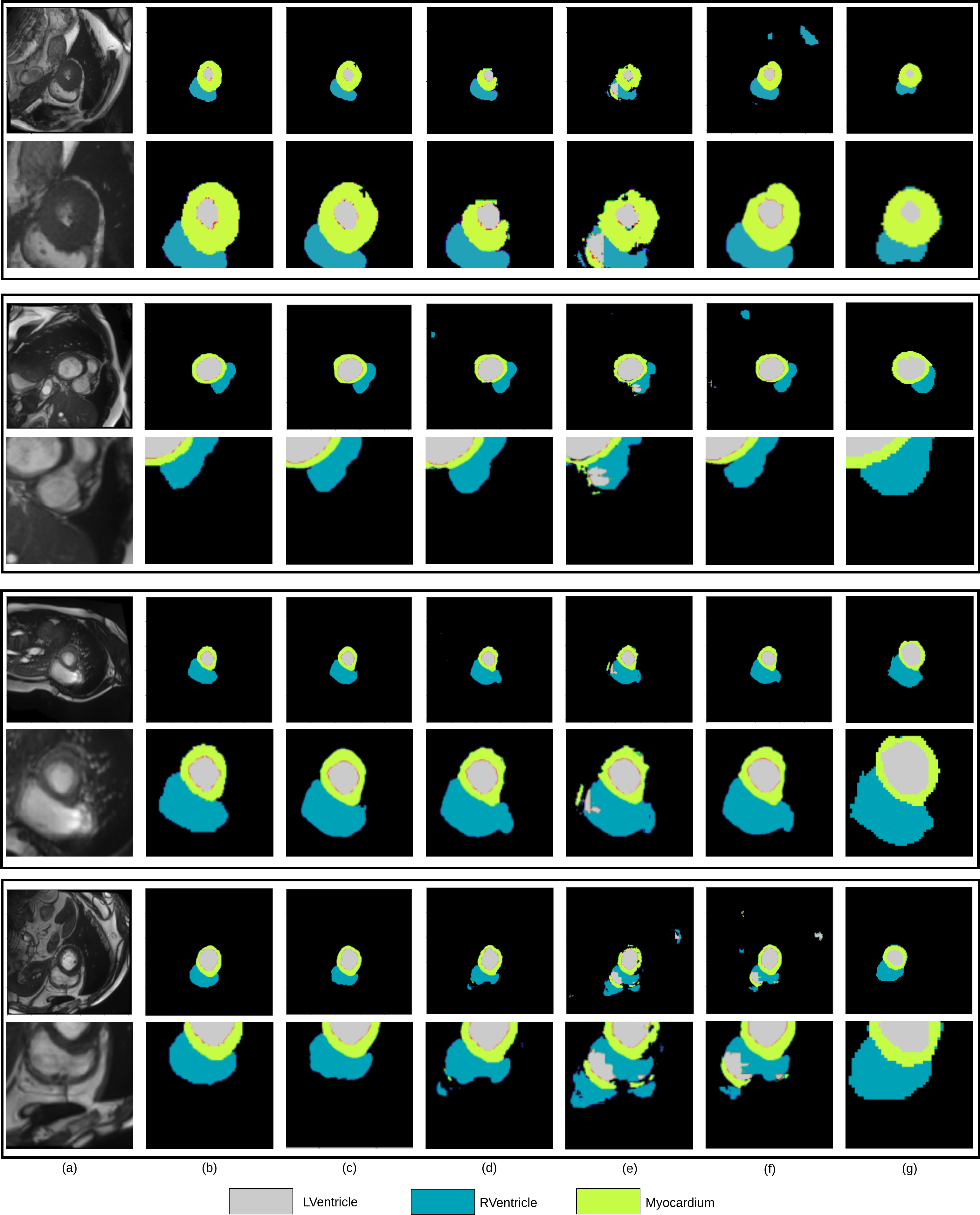}
    \caption{Comparative performance of {\it U-VixLSTM} and other baseline architectures, on the {\it ACDC} dataset, through sample segmentation maps. The first row in each block
    represents a sample CT slice. The second row in each block provides zoomed-in boxes to provide a magnified view of specific regions. The (a) input CT image, (b) corresponding ground truth, with the respective output from (c) {\it U-VixLSTM}, (d) Swin UNETR, (e) UNETR, (f) {\it V}-Net, and (g) TransAttUNet}
    \label{fig:acdc}
\end{figure*}

Table \ref{table:acdc} presents a comprehensive analysis of performance in different organs in the dataset {\it ACDC} concerning the {\it DSC} value. The average results for HD95 and IoU are tabulated for various cardiac organs. {\it U-VixLSTM} has outperformed baselines, attaining a {\it DSC} value of 86.8\%, 83.45\% and 91.04\% for the three distinct cardiac organs, respectively. The mean IoU is 77.7\%, exceeding the best baseline by approximately 6\%. The dotplot in Fig. \ref{fig:dot_acdc} provides a detailed analysis of the quantitative results. {\it U-VixLSTM} consistently displays dots with larger radii and pale color intensities across all the cardiac components in Fig. \ref{fig:dot_acdc}(a) and (b). This illustrates a balanced and high accuracy across the multiple interconnected cardiac structures. Other approaches, such as DS-TransUNet, Swin U-Mamba, TransUNet and Swin UNETR demonstrate either comparable or lower $HD95$ value, as evidenced from Fig. \ref{fig:dot_acdc}(c). However, their performance in terms of {\it DSC} and {\it IoU} is either significantly lower, as shown in Table \ref{table:acdc} and Fig. \ref{fig:dot_acdc}(a) and (b) or have a higher standard deviation concerning $HD95$ (Fig. \ref{fig:dot_acdc}(c)) compared to {\it U-VixLSTM}. This suggests {\it U-VixLSTM} achieves improved and consistent performance in both overlap-based and boundary-based metrics. The ViL block at the bottleneck of {\it U-VixLSTM} efficiently captures the global view of the cardiac anatomy. This intermediate output guides the feature reconstruction path to generate output that is accurate regarding boundary delineation and region overlap.

Fig. \ref{fig:acdc} shows sample segmentation maps generated by {\it U-VixLSTM} along with various baseline methods. The visual representation illustrates the higher-quality outputs generated by our proposed methods relative to established baselines. Instances of mis-segmentations and false positive pixels are evident in the sample maps from Swin UNETR, UNETR, and $V$-Net, as illustrated in Fig. \ref{fig:acdc}(d)-(g).

\begin{figure*}[t]
    \centering
    \includegraphics[width=0.5\linewidth]{figs/UVixLSTM_featuremaps.png}
    \caption{Visualization of (a) input image, (b) ground truth, and feature maps from (c) CNN and (d) ViL blocks of the feature extraction path.}
    \label{fig:feat}
\end{figure*}

Fig. \ref{fig:feat} illustrates sample feature maps obtained from the convolution and ViL modules within the feature extraction pathway. The maps visualize the representations acquired by the two modules of the proposed network in relation to the target structures, as specified in the ground truth. The shades of red represent the regions of strongest activation where the model allocates the majority of its attention, whereas the shades of dark blue signify the lowest activation levels. This visualization helps us observe the areas in the input image that are the most important to the model to define the target structure.

The enhanced performance of {\it U-VixLSTM} in various modalities is due to its effective capture of local and contextual information. The CNN blocks initially hierarchically extract fine-grained details, such as edges and local patterns. The ViL blocks in the bottleneck connect distant sections of intermediate feature maps, effectively capturing the general relationships and dependencies between various parts of target structures that have different shapes and sizes. This effectively constructs global contextual information. Incorporating skip connections that merge feature maps from the extraction path with the corresponding levels of the reconstruction path facilitates the localization of anatomical structures. Consequently, ViL enhance the feature extraction process by acquiring superior feature representations while maintaining lower computational costs relative to baseline methods. The gating mechanism of ViL selectively updates the memory matrix to store sharp transitions corresponding to the organ boundaries. This explains the superiority of {\it U-VixLSTM} in efficiently modelling the boundaries of diverse anatomical structures in comparison to Transformer-based models. Transformer computes the contextual embedding of each patch using a weighted average of all the image patches. Although being able to effectively capture the global context, the averaging step smoothens the fine-grained details necessary for capturing the boundaries.

\begin{figure}[t]
    \centering
    \includegraphics[width=0.7\linewidth]{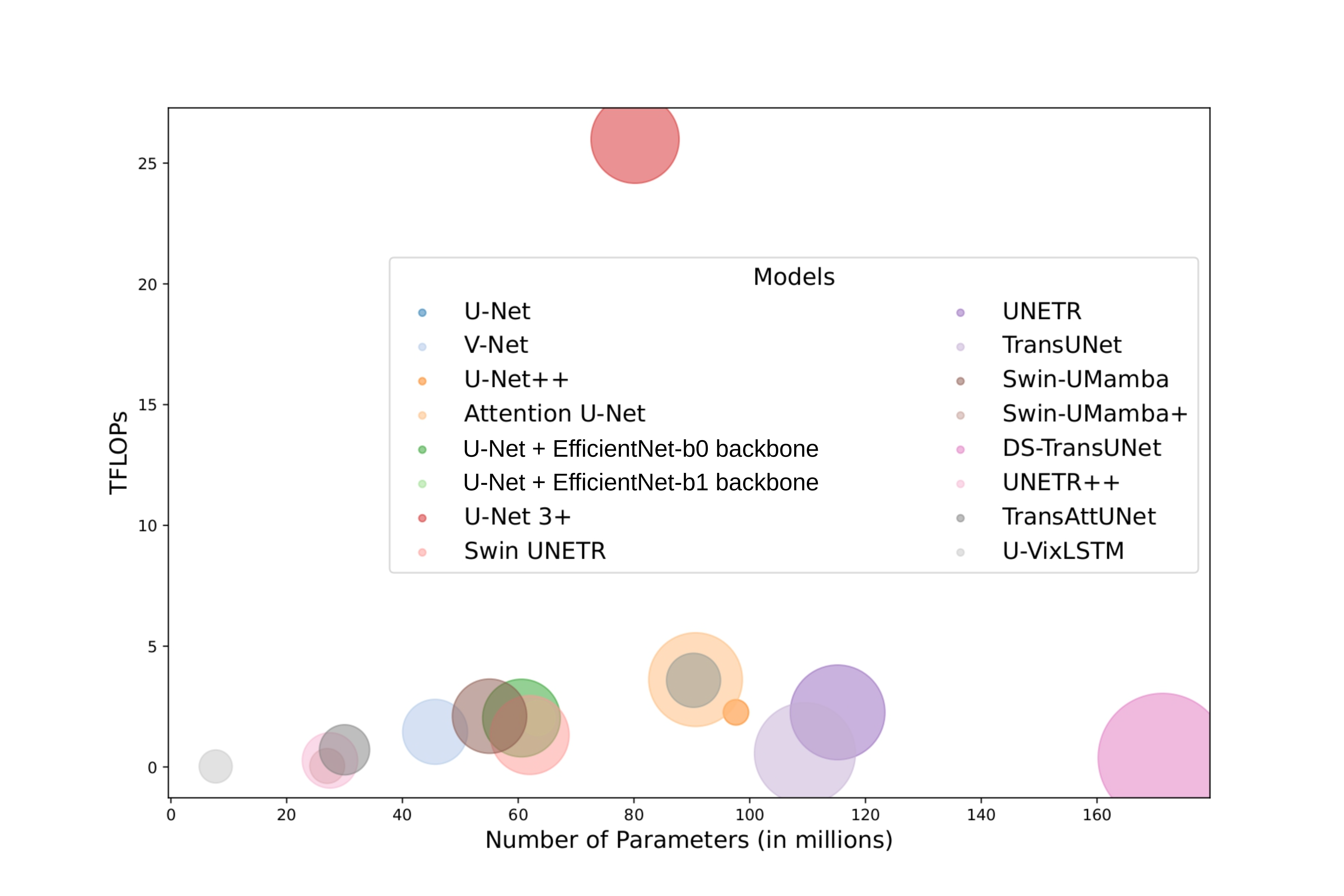}
    \caption{Comparison between state-of-the-art models  with respect to number of parameters (in millions), TFLOPs, and model size on disk (in MB). Bubble size is indicative of  model size.}
    \label{fig:params}
\end{figure}

Fig. \ref{fig:params} provides a graphical analysis of the parameter count, Tera FLoating point OPerations (TFLOPs), and model size on disk, of our {\it U-VixLSTM} compared to that of other baseline architectures under consideration. The analysis shows that {\it U-VixLSTM} has the lowest number of parameters and TFLOPs (floating point operations per second) compared to the other SOTA. This reiterates our claim about the superior computational efficiency of the proposed model. Hence, it demonstrates potential for deployment in resource-constrained environments.  

\section{Conclusions} \label{concl}
 
This research introduced the first integration of CNNs with ViL for image segmentation (specifically medical image segmentation), using the widely used $U$-shaped framework.  CNNs acquired intricate spatial and textural information from the input image. The ViL captured both global and temporal relationships inside the patches, derived from the feature volume produced by CNNs; thus learning a robust representation for the target structures. 

The superiority of {\it U-VixLSTM} can be attributed to various factors. Incorporation of CNNs with ViL allowed the model to effectively capture both local and global contextual information, which was essential to achieve precise segmentation. The exponential gating mechanism enables the model to efficiently retain longer contextual information necessary to precisely delineate target structures of variable sizes. Furthermore, the utilization of ViL improved the efficiency of the model in terms of the total number of parameters and FLOPs, compared to other hybrid CNN transformer methods. This has substantial implications for clinical practice by increasing the accessibility of our model in a scenario with limited resources. This can potentially lead to an expedited and effective identification of medical conditions, decision-making for treatment, and providing guidance during surgical procedures. Furthermore, the remarkable performance highlights the potential of ViL for effective medical image segmentation. This study lays the groundwork for further research and exploration in the application of ViL to this field.

\section*{Acknowledgement}

This research was supported by J. C. Bose National Fellowship grant JCB/2020/000033 of S. Mitra.
\bibliographystyle{ieeetr}
\bibliography{egbib}

\end{document}